\documentclass[conf]{new-aiaa} 
\usepackage[utf8]{inputenc}

\usepackage{graphicx}

\usepackage{amsmath}
\usepackage{dsfont}
\usepackage{hhline}
\usepackage{booktabs}
\usepackage{tabularx}
\usepackage{multirow}
\usepackage{boldline}
\usepackage{bm}
\usepackage{makecell}
\usepackage{indentfirst}
\usepackage{diagbox}

\usepackage{amsmath} 
\usepackage{amsfonts}

\usepackage{float} 

\usepackage{subcaption}

\usepackage{tikz}
\usetikzlibrary{circuits.ee.IEC}
\usetikzlibrary{shapes,arrows,calc,positioning}
\tikzstyle{bigblock} = [draw, fill=blue!20, rectangle, 
    minimum height=6em, minimum width=8em]
\tikzstyle{medblock} = [draw, fill=blue!20, rectangle, 
    minimum height=4em, minimum width=4em]    
\tikzstyle{mux} = [draw, fill=black!20, rectangle, 
    minimum height=5em, minimum width=0.1em]    
\tikzstyle{smallblock} = [draw, fill=blue!20, rectangle, 
    minimum height=2em, minimum width=3em]
    
\tikzstyle{data_block} = [draw, fill=green!20, rectangle, 
    minimum height=2em, minimum width=3em]
\tikzstyle{ops_block} = [draw, fill=blue!20, rectangle, 
    minimum height=2em, minimum width=3em]    
\tikzstyle{est_block} = [draw, fill=red!20, rectangle, 
    minimum height=2em, minimum width=3em]    
    
\tikzstyle{sum} = [draw, fill=blue!20, circle, node distance=1cm,minimum height=0.5cm]
\tikzstyle{signal} = [coordinate]
\tikzstyle{pinstyle} = [pin edge={to-,thin,black}]
\tikzstyle{block} = [draw, fill=blue!20, rectangle, 
    minimum height=3em, minimum width=9em]
\tikzstyle{blockS} = [draw, fill=blue!20, rectangle, 
    minimum height=3em, minimum width=4em]    
\tikzstyle{input} = [coordinate]
\tikzstyle{output} = [coordinate]

\tikzstyle{gain} = [draw, fill=blue!20, regular polygon, regular polygon sides=3, shape border rotate=30]
\tikzstyle{gain_vert} = [draw, fill=blue!20, regular polygon, regular polygon sides=3, shape border rotate=0]

\usetikzlibrary{matrix}

\usetikzlibrary{positioning}
\usetikzlibrary{math}

\usepackage{hyperref}
\usepackage{xcolor}
\hypersetup{
    colorlinks,
    linkcolor={blue!100!black},
    citecolor={blue!50!black},
    urlcolor={blue!80!black}
}

\newcommand{\bc}{\begin{center}}
\newcommand{\ec}{\end{center}}
\newcommand{\benum}{\begin{enumerate}}
\newcommand{\eenum}{\end{enumerate}}

\newcommand{\matl}{\left[ \begin{array}}
\newcommand{\matr}{\end{array} \right]}

\renewcommand{\matl}{\begin{bmatrix}}
\renewcommand{\matr}{\end{bmatrix}}

\newcommand{\matls}{\left[ \begin{smallmatrix}}
\newcommand{\matrs}{\end{smallmatrix} \right]}
\newcommand{\isdef}{\stackrel{\triangle}{=}}

\newcommand{\rmT}{{\rm T}}

\newcommand{\rma}{{\rm a}}

\newcommand{\rmd}{{\rm d}}

\newcommand{\rmp}{{\rm p}}

\newcommand{\BBR}{{\mathbb R}}

\usepackage{enumitem,amssymb}
\newlist{todolist}{itemize}{2}
\setlist[todolist]{label=$\square$}
\usepackage{pifont}

\usepackage[colorinlistoftodos,textwidth=1 in,textsize=footnotesize]{todonotes}
\DeclareMathOperator*{\argmin}{arg\,min}

\usepackage{algorithm}
\usepackage{algpseudocode}

\title{Control-Oriented Modeling for Adaptive and Decentralized Collaborative Payload Lifting}

\title{Modeling and Hierarchical Control of\\Collaborative Payload Lifting}

\author{
Anonno Singha Ray\footnote{Graduate Student, Department of Mechanical Engineering. \texttt{anonnor1@umbc.edu}.},
Alexander Dorsey\footnote{Graduate Student, Department of Mechanical Engineering. \texttt{adorsey4@umbc.edu}.},
and Ankit Goel\footnote{Assistant Professor, Department of Mechanical Engineering. \texttt{ankgoel@umbc.edu}.}
}

\affil{University of Maryland, Baltimore County, Baltimore, MD 21250}

\begin{document}

\maketitle


\begin{abstract}
Collaborative payload transportation using multiple aerial vehicles involves coupled dynamics arising from cable-mediated interactions between the payload and the agents.
This paper develops a constrained dynamical model and a hierarchical feedback controller for collaborative payload lifting using massless, inextensible cables.
By enforcing the cable-length constraints and their acceleration compatibility conditions, an explicit relationship between cable tensions, system states, and agent forces is derived.
Using this relationship, an outer-loop controller determines the desired cable tensions through constrained optimization, while an inner-loop controller computes the agent forces required to realize these tensions.
A transverse component of each agent's control force is used to regulate the angular separation between cables without affecting the desired cable tensions.
A decentralized force-realization approximation is also investigated by neglecting the inter-agent coupling terms.
Numerical simulations of a three-agent system demonstrate payload trajectory tracking and exact tension realization under the centralized controller, whereas the decentralized approximation introduces discrepancies between desired and realized tensions and increases payload tracking errors.
\end{abstract}

\section{Introduction}
\label{sec:introduction}

Cooperative transport of a payload by a team of aerial vehicles connected to it through cables allows the team to carry a load that a single vehicle cannot carry alone.
Applications include cooperative lifting by aerial vehicles \cite{michael2011cooperative,fink2011planning}, cable-suspended robotic manipulation, distributed logistics, construction in constrained environments, and payload deployment in hazardous or inaccessible regions \cite{maza2010multi}.
In these systems, the agents interact through cables that transmit forces to a shared payload, so the motion of each agent affects the forces acting on the payload and the other agents.
The cable tensions are not independent inputs, since they are determined by the payload dynamics, cable geometry, and motion of the agents.
The control problem is to command the agents so that the payload follows a desired trajectory while maintaining nonnegative cable tensions and suitable cable configurations.

Studies of cable-suspended transport differ in how the cables, payload, and agents are modeled.
Early work considered static equilibrium and quasi-static configurations \cite{michael2011cooperative,fink2011planning}, and subsequent work developed dynamic models for point-mass and rigid-body payloads \cite{sreenath2013geometric,lee2013geometric}.
Reduced models based on pendulum approximations or small-angle assumptions have also been investigated \cite{bisgaard2009modeling}.
Taut cables are commonly modeled as massless, inextensible connections, while flexible cables have been represented using lumped masses and spring-damper elements \cite{sharma2023decentralized,estevez2024review}.
Other studies consider rigid attachments between agents and payloads \cite{romadov2024control,culbertson2021decentralized} or rigid rods and links \cite{zeng2025decentralized}.
A broad review of modeling approaches for aerial transportation of suspended payloads is presented in \cite{estevez2024review}.

A variety of control strategies have been developed for cooperative payload transportation, including feedback control, formation control, adaptive control, and learning-based approaches \cite{sharma2023decentralized,zeng2025decentralized,culbertson2021decentralized,costantini2024cooperative}.
Both centralized \cite{romadov2024control,jimenez2022precise} and decentralized \cite{gassner2017dynamic,loianno2017cooperative} architectures have been investigated.
Recent studies have developed robust tension-based control for cooperative transportation using multiple aerial vehicles \cite{bacheti2026reconfigurable} and decentralized geometric control with
local estimation of payload motion and load sharing \cite{hajieghrary2026decentralized}.

Although these studies have developed models and controllers for cooperative payload transportation, an important consideration is the relationship between the desired cable tensions and the forces applied to the individual agents.
The cable tensions must satisfy the geometric constraints imposed by the cables and the coupled equations of motion.
Consequently, determining the agent forces required to realize prescribed tensions involves the dynamics of the entire payload-agent system.
Making this relationship explicit also provides a basis for investigating the effect of neglecting inter-agent coupling in the control law.

This paper develops a constrained dynamical model and a hierarchical feedback controller for cooperative payload transportation using multiple ideal cables.
The formulation follows from Newtonian mechanics and geometric compatibility constraints in the taut-cable regime.
Differentiating the constraints twice yields a linear system that determines the cable tensions from the system state and the forces applied to the agents.
Using this model, an outer-loop controller computes the desired cable tensions through constrained optimization, while an inner-loop controller determines the agent forces required to realize these tensions.
The remaining freedom in the agent forces is used to regulate cable geometry without changing the desired tensions.
A decentralized approximation is also considered by neglecting the inter-agent coupling terms in the force-realization controller.

The contributions of this paper are as follows.
\begin{itemize}
    \item A constrained dynamical model of cooperative payload lifting with an explicit linear system relating cable tensions to the system state and agent forces.
    \item A hierarchical feedback controller that combines constrained tension allocation, exact force realization under the assumed model, and cable geometry regulation, together with a decentralized force-realization approximation obtained by neglecting inter-agent coupling terms.
    \item Closed-loop numerical simulations of a three-agent system in three dimensions to evaluate payload trajectory tracking, cable tensions, and the effects of decentralized force realization.
\end{itemize}

The remainder of this paper is organized as follows.
Section~\ref{sec:model} develops the constrained dynamics and cable-tension equations.
Section~\ref{sec:tension_realizing_control} presents the hierarchical feedback controller and decentralized force-realization approximation.
Section~\ref{sec:numerical_simulation} presents closed-loop numerical results for a three-agent configuration.
Section~\ref{sec:conclusions} summarizes the paper.

\section{Collaborative Payload Lifting Model}
\label{sec:model}
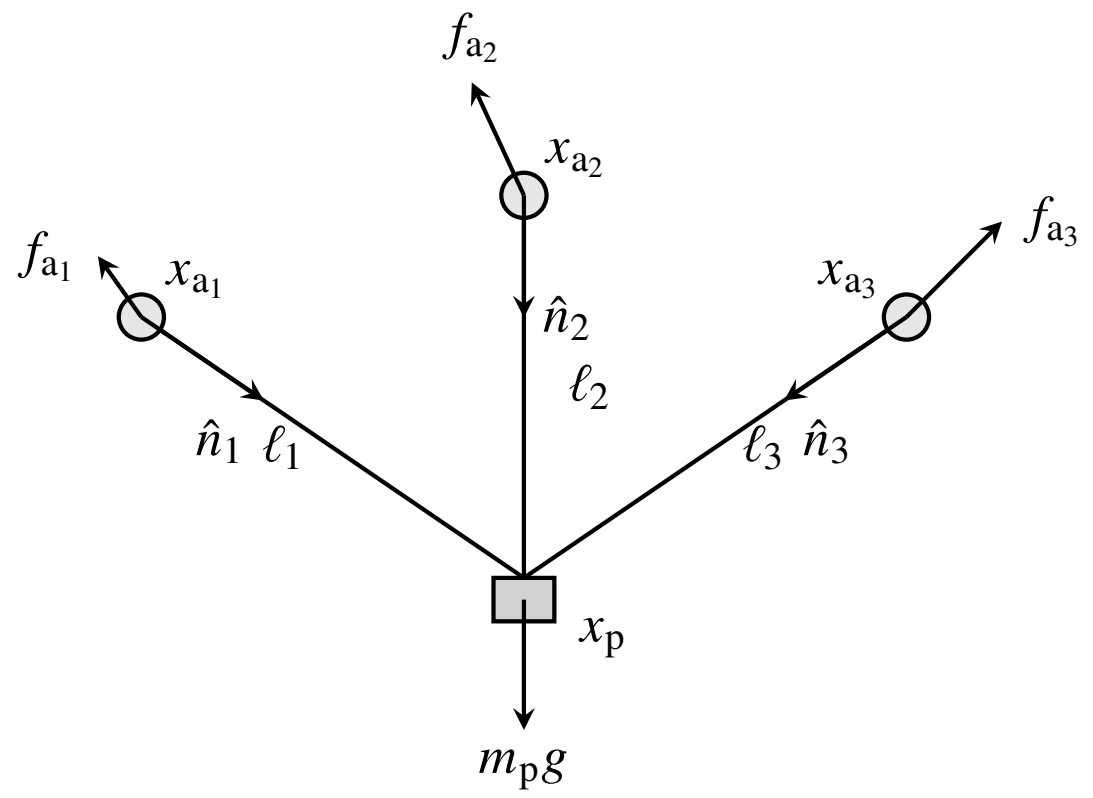
\begin{figure}[h]
    \centering
    \begin{tikzpicture}[scale=1.15, >=stealth]

        \coordinate (P)  at (0,0);
        \coordinate (A1) at (-2.2,1.5);
        \coordinate (A2) at (0,2.2);
        \coordinate (A3) at (2.2,1.5);

        \draw[thick] (A1) -- (P);
        \draw[thick] (A2) -- (P);
        \draw[thick] (A3) -- (P);

        \filldraw[fill=gray!20, draw=black, thick] (A1) circle (0.13);
        \filldraw[fill=gray!20, draw=black, thick] (A2) circle (0.13);
        \filldraw[fill=gray!20, draw=black, thick] (A3) circle (0.13);

        \draw[fill=gray!35, draw=black, thick]
        ($(P)+(-0.175,0)$) rectangle ($(P)+(0.175,-0.25)$);
    
        \coordinate (Pc) at ($(P)+(0,-0.125)$);

        \node[above left, xshift=2em]  at (A1) {$x_{{\rm a}_1}$};
        \node[above, xshift=1em]       at (A2) {$x_{{\rm a}_2}$};
        \node[above right, xshift=-2em] at (A3) {$x_{{\rm a}_3}$};
        \node[below, xshift=1.5em]       at (Pc) {$x_{\rm p}$};

        \node[left]  at (-1.15,0.75) {$\ell_1$};
        \node[right] at (0.15,1.1) {$\ell_2$};
        \node[right] at (1.15,0.75) {$\ell_3$};

        \draw[->, thick] (A1) -- ($(A1)!0.32!(P)$);
        \draw[->, thick] (A2) -- ($(A2)!0.32!(P)$);
        \draw[->, thick] (A3) -- ($(A3)!0.32!(P)$);

        \node[below left]  at ($(A1)!0.32!(P)$) {$\hat n_1$};
        \node[right]       at ($(A2)!0.32!(P)$) {$\hat n_2$};
        \node[below right] at ($(A3)!0.32!(P)$) {$\hat n_3$};

        \draw[->, thick] (A1) -- ++(-0.25,0.35) node[left] {$f_{{\rm a}_1}$};
        \draw[->, thick] (A2) -- ++(-.3,0.65) node[above] {$f_{{\rm a}_2}$};
        \draw[->, thick] (A3) -- ++(0.55,0.55) node[right] {$f_{{\rm a}_3}$};

        \draw[->, thick] (Pc) -- ++(0,-0.75) node[below] {$m_{\rm p}g$};


    \end{tikzpicture}
    \caption{Planar three-agent collaborative lifting configuration. The payload position is denoted by $x_{\rm p}$, the actuating-agent positions are denoted by $x_{{\rm a}_i}$, and the payload is connected to each agent through an ideal string of fixed length $\ell_i$.}
    \label{fig:planar_lifting_geometry}
\end{figure}

This section develops the equations of motion of a payload suspended from multiple agents through ideal cables.
The geometric constraints imposed by the cables are used to derive an explicit expression for the cable tensions in terms of the system state and the forces applied by the agents.
Figure~\ref{fig:planar_lifting_geometry} illustrates the system configuration for three agents, while the development applies to an arbitrary number of agents in three dimensions.

\subsection{System Description and Assumptions}
Consider a payload of mass $m_\rmp>0$ suspended by $n$ ideal cables from $n$ agents.
Let $x_\rmp(t)\in\BBR^3$ denote the payload position, and let $x_{\rma_i}(t)\in\BBR^3$ and $m_{\rma_i}>0$ denote the position and mass of agent $i$, for $i=1,\ldots,n$.
Let $f_{\rma_i}(t)\in\BBR^3$ denote the control force applied to agent $i$.
The payload and agents are modeled as point masses, and the control forces are assumed to be directly realizable.
Consequently, attitude dynamics, actuator dynamics, and aerodynamic effects are not modeled.
The cables are massless, inextensible, straight when taut, and capable of transmitting only tensile forces.
Let $\ell_i>0$ denote the fixed length of cable $i$, and let $g\in\BBR^3$ denote the gravitational acceleration vector.
The model describes the motion while every cable is taut, and a cable that becomes slack leaves the regime of the model.

\subsection{Geometric Constraints}
For $i=1,\ldots,n$, the inextensibility of the cables imposes the holonomic constraints
\begin{align}
    \|x_\rmp(t)-x_{\rma_i}(t)\|
    =
    \ell_i.
    \label{eq:string_constraints}
\end{align}
The constraints and their first time derivative hold at $t=0$, that is, $\|x_\rmp(0)-x_{\rma_i}(0)\|=\ell_i$ and $(x_\rmp(0)-x_{\rma_i}(0))^T(\dot x_\rmp(0)-\dot x_{\rma_i}(0))=0$.
Define the unit vector along cable $i$ from agent $i$ to the payload as
\begin{align}
    \hat n_i
    \isdef
    \frac{x_\rmp-x_{\rma_i}}{\|x_\rmp-x_{\rma_i}\|}
    =
    \frac{x_\rmp-x_{\rma_i}}{\ell_i},
\end{align}
and let $N(\xi)\isdef\begin{bmatrix}\hat n_1&\cdots&\hat n_n\end{bmatrix}\in\BBR^{3\times n}$ denote the matrix of cable directions and 
the state vector
\begin{align}
    \xi
    \isdef
    \begin{bmatrix}
        x_\rmp^T & x_{\rma_1}^T & \cdots & x_{\rma_n}^T &
        \dot x_\rmp^T & \dot x_{\rma_1}^T & \cdots & \dot x_{\rma_n}^T
    \end{bmatrix}^T
    \in\BBR^{6(n+1)}.
\end{align}
Let $T_i(t)$ denote the tension in cable $i$.
Since the cables cannot transmit compression, $T_i(t)\ge0$ for $i=1,\ldots,n$.

\subsection{Equations of Motion}
The payload dynamics satisfy
\begin{align}
    m_\rmp\ddot x_\rmp
    =
    m_\rmp g
    -
    \sum_{i=1}^n T_i\hat n_i,
    \label{eq:payload_dynamics}
\end{align}
where the sum equals $N T$ with $T\isdef\begin{bmatrix}T_1&\cdots&T_n\end{bmatrix}^T$.
The dynamics of agent $i$ are
\begin{align}
    m_{\rma_i}\ddot x_{\rma_i}
    =
    m_{\rma_i}g
    +
    f_{\rma_i}
    +
    T_i\hat n_i.
    \label{eq:agent_dynamics}
\end{align}
Equations \eqref{eq:payload_dynamics}--\eqref{eq:agent_dynamics} define a constrained multibody system coupled through the cable tensions.

\subsection{Constraint-Compatible Tension Equations}
Differentiating the constraint \eqref{eq:string_constraints} yields
$(x_\rmp-x_{\rma_i})^T(\dot x_\rmp-\dot x_{\rma_i})=0$.
Differentiating again gives
\begin{align}
    (x_\rmp-x_{\rma_i})^T
    (\ddot x_\rmp-\ddot x_{\rma_i})
    +
    \|\dot x_\rmp-\dot x_{\rma_i}\|^2
    =
    0.
    \label{eq:acc_constraint}
\end{align}
Using $x_\rmp-x_{\rma_i}=\ell_i\hat n_i$ and substituting \eqref{eq:payload_dynamics}--\eqref{eq:agent_dynamics} into \eqref{eq:acc_constraint} yields
\begin{align}
    \ell_i
    \hat n_i^T
    \Bigg[
        -\frac{1}{m_{\rma_i}}f_{\rma_i}
        -
        \sum_{j=1}^n
        \frac{1}{m_\rmp}
        T_j \hat n_j
        -
        \frac{T_i}{m_{\rma_i}}
        \hat n_i
    \Bigg]
    +
    \|\dot x_\rmp-\dot x_{\rma_i}\|^2
    =
    0.
\end{align}

Collecting the tension terms yields
\begin{align}
    \sum_{j=1}^n M_{ij}(\xi)T_j
    =
    \rho_i(\xi)
    +
    H_i(\xi)f_{\rma_i},
\end{align}
where
\begin{align}
    M_{ij}(\xi)
    &\isdef
    \ell_i
    \left(
        \frac{1}{m_\rmp}
        \hat n_i^T\hat n_j
        +
        \delta_{ij}\frac{1}{m_{\rma_i}}
    \right),
    \quad 
    \rho_i(\xi)
    \isdef
    \|\dot x_\rmp-\dot x_{\rma_i}\|^2,
    \quad
    H_i(\xi)
    \isdef
    -\frac{\ell_i}{m_{\rma_i}}\hat n_i^T.
\end{align}
Defining
$u
    \isdef
    \begin{bmatrix}
        f_{\rma_1}^T & \cdots & f_{\rma_n}^T
    \end{bmatrix}^T$
and 
$H(\xi)
    \isdef
    \operatorname{diag}\big(H_1(\xi),\ldots,H_n(\xi)\big),$
the tensions satisfy
\begin{align}
    M(\xi)T
    =
    \rho(\xi)
    +
    H(\xi)u.
    \label{eq:tension_system}
\end{align}

The matrix $M(\xi)$ is nonsingular for every $\xi$.
To see this, write
$M(\xi)=\operatorname{diag}(\ell_1,\ldots,\ell_n)P(\xi)$,
where
\begin{align}
    P(\xi)
    \isdef
    \frac{1}{m_\rmp}N^TN
    +
    \operatorname{diag}
    \left(
        \frac{1}{m_{\rma_1}},
        \ldots,
        \frac{1}{m_{\rma_n}}
    \right).
\end{align}
The matrix $N^TN$ is positive semidefinite and the diagonal matrix is positive definite, so $P(\xi)$ is positive definite and $M(\xi)$ is the product of two nonsingular matrices.
The rank of $N(\xi)$ depends on the cable geometry and determines the payload forces that the cables can produce, while $M(\xi)$ remains nonsingular when $N(\xi)$ loses rank.
The tensions are therefore
\begin{align}
    T
    =
    M(\xi)^{-1}
    \big(
        \rho(\xi)+H(\xi)u
    \big).
    \label{eq:tension_solution}
\end{align}
Equation \eqref{eq:tension_solution} shows that the cable tensions are not independent inputs and are determined by the state and the agent forces.
The model describes the motion on the set of states and inputs for which every component of $T$ in \eqref{eq:tension_solution} is nonnegative.

\section{Tension-Realizing Payload Feedback Control}
\label{sec:tension_realizing_control}
This section develops a feedback controller that makes the payload position $x_\rmp$ track a desired trajectory $x_{\rmp,\rmd}(t)$ using the agent forces $f_{\rma_i}$.
The payload is not actuated directly, and its motion is produced by the cable tensions that the agent forces induce through \eqref{eq:tension_solution}.
The controller therefore has two stages, in which an outer loop selects the cable tensions that give the payload the desired acceleration and an inner loop selects the agent forces that produce those tensions.
A component of each agent force transverse to its cable does not affect the tensions, and this component is used to regulate the angular separation between the cables.

\subsection{Outer-Loop Tension Allocation}
\label{subsec:outer_loop_tension_allocation}
Let $e_\rmp\isdef x_\rmp-x_{\rmp,\rmd}$ and $\dot e_\rmp\isdef\dot x_\rmp-\dot x_{\rmp,\rmd}$, and choose the desired payload acceleration as
\begin{align}
    a_{\rmp,\rmd}
    \isdef
    \ddot x_{\rmp,\rmd}
    -
    K_\rmd\dot e_\rmp
    -
    K_\rmp e_\rmp,
    \label{eq:tension_control_desired_payload_acceleration}
\end{align}
where $K_\rmp$ and $K_\rmd$ are symmetric positive definite matrices.
By \eqref{eq:payload_dynamics}, the payload dynamics are $m_\rmp\ddot x_\rmp=m_\rmp g-N(\xi)T$, so a tension vector $T_\rmd$ that gives $\ddot x_\rmp=a_{\rmp,\rmd}$ satisfies
\begin{align}
    N(\xi)T_\rmd
    =
    m_\rmp
    \big(
        g-a_{\rmp,\rmd}
    \big).
    \label{eq:tension_control_desired_tension_allocation}
\end{align}
Let $r\isdef N(\xi)T_\rmd-m_\rmp(g-a_{\rmp,\rmd})$ denote the residual of \eqref{eq:tension_control_desired_tension_allocation}.
If the tension $T=T_\rmd$ is realized, then $\ddot x_\rmp=a_{\rmp,\rmd}-r/m_\rmp$, and the tracking error satisfies
\begin{align}
    \ddot e_\rmp
    +
    K_\rmd\dot e_\rmp
    +
    K_\rmp e_\rmp
    =
    -\frac{1}{m_\rmp}r.
    \label{eq:tension_control_error_dynamics}
\end{align}
For $r=0$, the function $V=\tfrac12\dot e_\rmp^T\dot e_\rmp+\tfrac12e_\rmp^TK_\rmp e_\rmp$ satisfies $\dot V=-\dot e_\rmp^TK_\rmd\dot e_\rmp$, so $e_\rmp$ converges to zero.

The matrix $N(\xi)$ has $n$ columns and three rows, so \eqref{eq:tension_control_desired_tension_allocation} may have no solution with positive tensions or may have many.
Cables must remain taut, and the tension is therefore required to satisfy $T\geq T_0$, where $T_0\in\BBR^n$ has positive entries.
The desired tension is obtained from two quadratic programs.
The first program minimizes the residual,
\begin{align}
    T^\ast
    =
    \argmin_{T\geq T_0}
    \big\|
        N(\xi)T
        -
        m_\rmp
        \big(
            g-a_{\rmp,\rmd}
        \big)
    \big\|^2,
    \label{eq:tension_control_qp_residual}
\end{align}
and the second program selects the minimizer closest to the lower bound,
\begin{align}
    T_\rmd
    =
    \argmin_{T\geq T_0}
    \|T-T_0\|^2
    \quad
    \text{subject to}
    \quad
    N(\xi)T=N(\xi)T^\ast.
    \label{eq:tension_control_qp_selection}
\end{align}
All minimizers of \eqref{eq:tension_control_qp_residual} share the same product $N(\xi)T$, because the objective is strictly convex in $N(\xi)T$.
The feasible set of \eqref{eq:tension_control_qp_selection} is therefore the set of minimizers, which is nonempty, closed, and convex, and the strictly convex objective gives a unique $T_\rmd$.
The residual is $r=N(\xi)T^\ast-m_\rmp(g-a_{\rmp,\rmd})$, which is zero if and only if some $T\geq T_0$ satisfies \eqref{eq:tension_control_desired_tension_allocation}.

\subsection{Inner-Loop Force Realization}
\label{subsec:inner_loop_force_realization}
By \eqref{eq:tension_system}, the tension $T=T_\rmd$ is realized if and only if
\begin{align}
    H(\xi)u
    =
    M(\xi)T_\rmd
    -
    \rho(\xi).
    \label{eq:tension_control_realization_equation}
\end{align}
The rows of $H(\xi)$ act on different agent forces and each is nonzero, so $H(\xi)H(\xi)^T=\operatorname{diag}(\ell_i^2/m_{\rma_i}^2)$.
The matrix $H(\xi)$ therefore has full row rank, and \eqref{eq:tension_control_realization_equation} has a solution for every geometry.
The minimum-norm solution is $u=H^\dagger(\xi)(M(\xi)T_\rmd-\rho(\xi))$, and agent $i$ applies the force
\begin{align}
    f_{\rma_i}^{\rm r}
    =
    -
    \frac{m_{\rma_i}}{\ell_i}
    e_i^\rmT
    \big(
        M(\xi)T_\rmd
        -
        \rho(\xi)
    \big)
    \hat n_i.
    \label{eq:tension_control_explicit_agent_force}
\end{align}
Since $\hat n_i^T\hat n_i=1$, this force satisfies $-(\ell_i/m_{\rma_i})\hat n_i^Tf_{\rma_i}^{\rm r}=e_i^T(M(\xi)T_\rmd-\rho(\xi))$, which is the $i$th row of \eqref{eq:tension_control_realization_equation}.

With the force \eqref{eq:tension_control_explicit_agent_force}, the tension given by \eqref{eq:tension_solution} equals $T_\rmd$ at every time.
Since $T_\rmd\geq T_0>0$, every cable remains taut along the closed-loop solution, and the model of Section~\ref{sec:model} applies, whether or not the residual $r$ is zero.
The residual enters only through \eqref{eq:tension_control_error_dynamics} and degrades payload tracking when no $T\geq T_0$ satisfies \eqref{eq:tension_control_desired_tension_allocation}.

\subsection{Decentralized Force Realization}
\label{subsec:decentralized_force_realization}
The force realization controller \eqref{eq:tension_control_explicit_agent_force} depends on the states and desired tensions of all agents.
Expanding the expression for the control force of agent $i$ gives
\begin{align}
    f_{\rma_i}^{\rm r}
    &=
    -\frac{m_{\rma_i}}{\ell_i}
    \left(
        \sum_{j=1}^n M_{ij}(\xi)T_{\rmd,j}
        -
        \rho_i(\xi)
    \right)\hat n_i.
\end{align}
Separating the terms involving agent $i$ from those involving the other agents yields
\begin{align}
    f_{\rma_i}^{\rm r}
    &=
    f_{\rma_i}^{\rm local}
    +
    f_{\rma_i}^{\rm coupling},
\end{align}
where
\begin{align}
    f_{\rma_i}^{\rm local}
    &\isdef
    -\frac{m_{\rma_i}}{\ell_i}
    \left(
        M_{ii}(\xi)T_{\rmd,i}
        -
        \rho_i(\xi)
    \right)\hat n_i,
    \quad 
    f_{\rma_i}^{\rm coupling}
    \isdef
    -\frac{m_{\rma_i}}{\ell_i}
    \sum_{j\neq i}
    M_{ij}(\xi)T_{\rmd,j}\hat n_i.
    \label{eq:decentralized_force_realization_a}
\end{align}
The decentralized force realization is obtained by neglecting $f_{\rma_i}^{\rm coupling}$, so that
\begin{align}
    f_{\rma_i}^{\rm r}
    =
    f_{\rma_i}^{\rm local}.
    \label{eq:decentralized_force_realization}
\end{align}
Unlike the centralized controller, this approximation does not generally guarantee that the realized cable tensions equal the desired tensions.

\subsection{Angular Separation Between Cables}
\label{subsec:angular_separation}
The force of agent $i$ can include a component perpendicular to its cable without changing the tension.
Let $P_i\isdef I-\hat n_i\hat n_i^T$, and let
\begin{align}
    f_{\rma_i}
    =
    f_{\rma_i}^{\rm r}
    +
    P_iw_i,
    \qquad
    i=1,\ldots,n.
    \label{eq:tension_control_total_force}
\end{align}
Since $\hat n_i^TP_i=0$, the second term does not enter the $i$th row of \eqref{eq:tension_control_realization_equation}, so $T=T_\rmd$ for every choice of $w_i$.

The cable directions must remain separated to avoid a configuration in which the cables are parallel.
Let $\theta_{\min}>0$ denote the minimum allowable angle between any pair of cables.
Since $\hat n_i^T\hat n_j=\cos\theta_{ij}$, the condition on the angle is, for $i\neq j$,
\begin{align}
    c_{ij}(\xi)
    \isdef
    \hat n_i^T\hat n_j
    \leq
    \cos\theta_{\min},
    \label{eq:tension_control_angular_separation_constraint}
\end{align}
and its violation is measured by
\begin{align}
    \Delta_{ij}(\xi)
    \isdef
    \max
    \left\{
        0,
        c_{ij}(\xi)-\cos\theta_{\min}
    \right\}.
    \label{eq:tension_control_angular_separation_violation}
\end{align}
The gradient of $c_{ij}$ with respect to the agent position is
\begin{align}
    \frac{\partial c_{ij}}{\partial x_{\rma_i}}
    =
    -
    \frac{1}{\ell_i}
    P_i\hat n_j,
    \label{eq:tension_control_alignment_gradient}
\end{align}
so a transverse force along $P_i\hat n_j$ accelerates agent $i$ in a direction that decreases $c_{ij}$.
The transverse input is
\begin{align}
    w_i
    =
    -
    m_{\rma_i}P_ig
    +
    k_\theta
    \sum_{j\neq i}
    \Delta_{ij}(\xi)
    P_i\hat n_j
    -
    k_{\theta,\rmd}
    P_i\dot x_{\rma_i},
    \label{eq:tension_control_geometry_regulation_input}
\end{align}
where $k_\theta>0$ and $k_{\theta,\rmd}\geq0$ are gains.
The first term cancels the component of the agent weight transverse to the cable, the second term accelerates the agent away from cables that are nearly aligned with its own, and the third term damps the transverse velocity of the agent.

\subsection{Transverse Dynamics of the Cable Directions}
\label{subsec:transverse_dynamics}
The effect of \eqref{eq:tension_control_geometry_regulation_input} on the cable directions follows from the second derivative of $x_\rmp-x_{\rma_i}=\ell_i \hat n_i$, which gives $\ell_i\ddot{\hat n}_i=\ddot x_\rmp-\ddot x_{\rma_i}$.
Since $P_i\hat n_i=0$ and $P_if_{\rma_i}^{\rm r}=0$, the agent dynamics \eqref{eq:agent_dynamics} give $P_i\ddot x_{\rma_i}=P_ig+P_iw_i/m_{\rma_i}$.
Applying $P_i$ to the relative acceleration gives
\begin{align}
    \ell_iP_i\ddot{\hat n}_i
    =
    P_i
    \big(
        \ddot x_\rmp-g
    \big)
    -
    \frac{1}{m_{\rma_i}}P_iw_i.
    \label{eq:tension_control_cable_direction_dynamics}
\end{align}
Consider a payload at rest, so that $\dot x_\rmp=0$ and $\ddot x_\rmp=0$.
Without the first term of \eqref{eq:tension_control_geometry_regulation_input}, the right side of \eqref{eq:tension_control_cable_direction_dynamics} contains the forcing $-P_ig$.
The cable direction $\hat n_i$ can be stationary only if $P_ig=0$, that is, only if cable $i$ is parallel to $g$.
If this holds for all cables, the cables are parallel to each other and \eqref{eq:tension_control_angular_separation_constraint} is violated.
With the first term of \eqref{eq:tension_control_geometry_regulation_input}, the gravity terms cancel and
\begin{align}
    \ell_iP_i\ddot{\hat n}_i
    =
    -
    \frac{1}{m_{\rma_i}}
    \left(
        k_\theta
        \sum_{j\neq i}
        \Delta_{ij}P_i\hat n_j
        -
        k_{\theta,\rmd}P_i\dot x_{\rma_i}
    \right).
    \label{eq:tension_control_cable_direction_regulated}
\end{align}
For a payload at rest, $\dot x_{\rma_i}=-\ell_i\dot{\hat n}_i$ and $\dot{\hat n}_i^T\hat n_i=0$.
If $\Delta_{ij}=0$ for all $j\neq i$, then \eqref{eq:tension_control_cable_direction_regulated} gives $P_i\ddot{\hat n}_i=-(k_{\theta,\rmd}/m_{\rma_i})\dot{\hat n}_i$.
Since $\dot{\hat n}_i^T\hat n_i=0$, the derivative of $\tfrac12\|\dot{\hat n}_i\|^2$ equals $\dot{\hat n}_i^TP_i\ddot{\hat n}_i$, so
\begin{align}
    \|\dot{\hat n}_i(t)\|
    =
    \|\dot{\hat n}_i(t_0)\|
    e^{-k_{\theta,\rmd}(t-t_0)/m_{\rma_i}}.
\end{align}
The angular rate of each cable therefore decays exponentially while the payload is at rest and the angular separation constraint is satisfied.
When the payload moves, the damping term acts on the absolute transverse velocity of the agent, which includes the velocity of the payload, and this statement does not apply.
The regulation is not a forward invariance guarantee for \eqref{eq:tension_control_angular_separation_constraint}.

\section{Numerical Simulation}
\label{sec:numerical_simulation}
This section presents numerical simulations of the collaborative payload lifting system with three agents.
Two control scenarios are considered.
The first uses the full state of the payload-agent system to compute the control force applied to each agent.
The second uses the decentralized force realization developed in Section~\ref{subsec:decentralized_force_realization}, in which the inter-agent coupling terms are neglected.
Both scenarios use the same system parameters, initial conditions, and desired payload trajectory.

\subsection{Simulation Setup}
Consider a payload of mass $m_\rmp=5\,\mathrm{kg}$  suspended by three agents, each having mass $m_{\rma_i}=2\,\mathrm{kg}$ and cable length $\ell_i=1.5\,\mathrm{m}$, for $i=1,2,3$.
The gravitational acceleration vector is $g=[0\;\;0\;\;-9.8]^T\,\mathrm{m/s^2}$.
The outer-loop feedback gains are
\begin{align}
    K_\rmp &= \operatorname{diag}(80,80,80),\\
    K_\rmd &= \operatorname{diag}(60,60,60),
    \label{eq:simulation_tracking_gains}
\end{align}
and the desired cable tensions satisfy $T_{\rmd,i}\geq 10\,\mathrm{N}$ for $i=1,2,3$.
The transverse damping gain is $k_{\theta,\rmd}=1$.

The initial payload and agent positions are chosen as
\begin{align}
    x_\rmp(0)
    &=
    \begin{bmatrix}
        0 & 0 & 0.5
    \end{bmatrix}^T
    \,\mathrm{m},\\
    x_{\rma_1}(0)
    &=
    \begin{bmatrix}
        -0.968 & 0.559 & 1.618
    \end{bmatrix}^T
    \,\mathrm{m},\\
    x_{\rma_2}(0)
    &=
    \begin{bmatrix}
        0 & -1.118 & 1.618
    \end{bmatrix}^T
    \,\mathrm{m},\\
    x_{\rma_3}(0)
    &=
    \begin{bmatrix}
        0.968 & 0.559 & 1.618
    \end{bmatrix}^T
    \,\mathrm{m}.
    \label{eq:simulation_initial_agent_positions}
\end{align}
These positions are chosen arbitrarily to satisfy the cable-length constraints
$\|x_\rmp(0)-x_{\rma_i}(0)\|=\ell_i$, for $i=1,2,3$.
The initial velocities of the payload and all agents are zero.

The desired payload trajectory is
\begin{align}
    x_{\rmp,\rmd}(t)
    =
    \begin{bmatrix}
        0.5\cos(\omega t)\\
        0.7\sin(\omega t)\\
        1+0.1\sin(\omega t)
    \end{bmatrix}
    \,\mathrm{m},
    \label{eq:simulation_reference_trajectory}
\end{align}
where $\omega=0.1\,\mathrm{rad/s}.$
The desired velocity and acceleration are obtained by
differentiating \eqref{eq:simulation_reference_trajectory}.
Both simulations are performed over $100\,\mathrm{s}$.

\subsection{Ideal Force Realization}
\label{subsec:sim_ideal}
In the first scenario, the control force applied to each agent
is computed using the full state $\xi$ and the centralized
force realization developed in
Section~\ref{subsec:inner_loop_force_realization}.
The desired cable tensions are determined by the outer-loop
tension allocation, and the agent forces are computed to
realize these tensions under the assumed model.
The following results evaluate payload trajectory tracking,
agent motion, control forces, and cable tensions.

\begin{figure}[h]
    \centering
    \includegraphics[width=.5\linewidth]{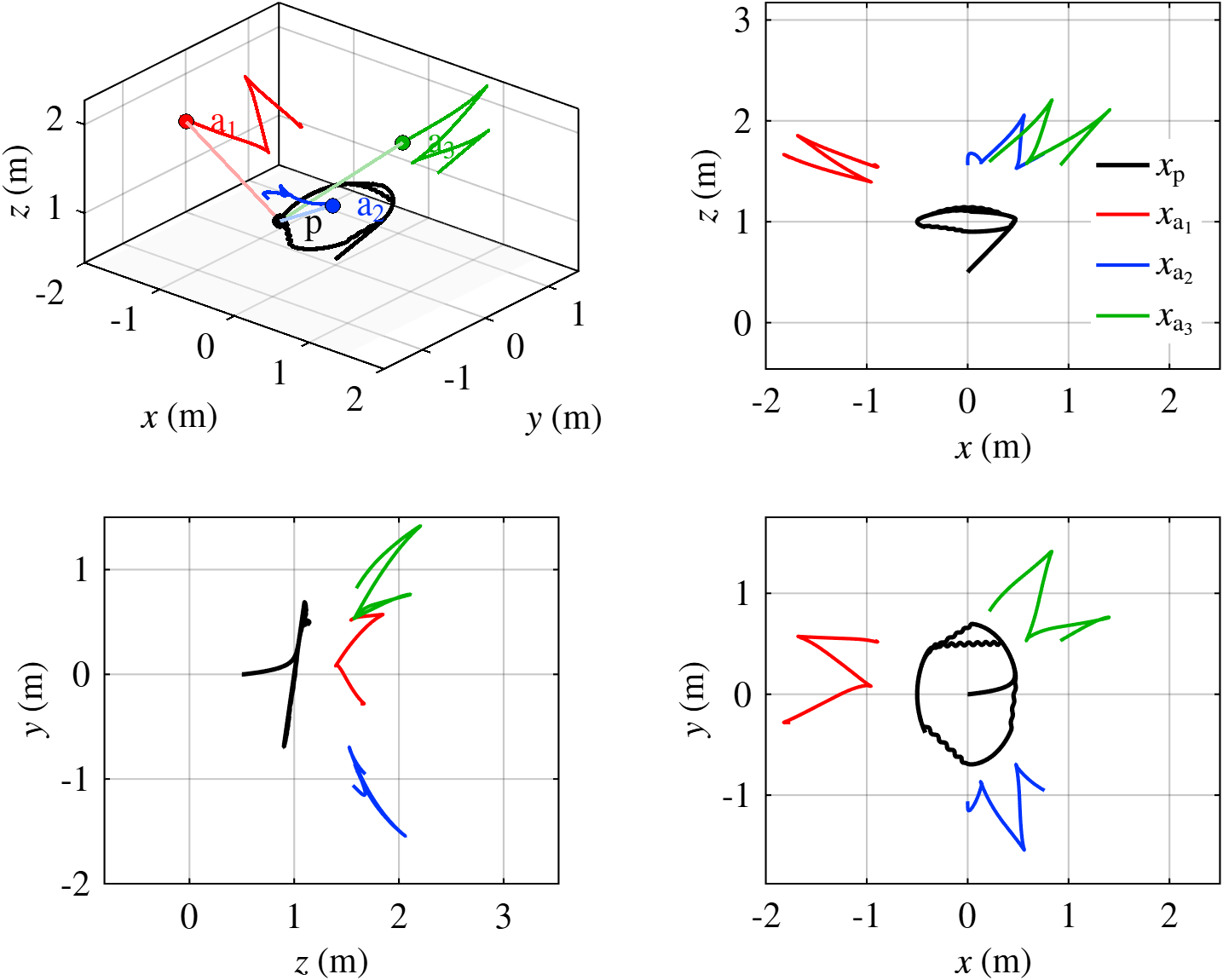}
    \caption{Ideal force realization. Trajectories of the payload $P$ and the agents $a_1$, $a_2$, and
    $a_3$ in three dimensions (top left) and their projections onto the
    $xz$ (top right), $zy$ (bottom left), and $xy$ (bottom right)
    planes. The payload trajectory $x_{\mathrm{p}}$ is shown in solid black, and the agent trajectories  $x_{\mathrm{a}_1}$,  $x_{\mathrm{a}_2}$, and  $x_{\mathrm{a}_3}$  are shown in red, blue, and green,
    respectively.}
    \label{fig:closed_loop_trajectories}
\end{figure}

Figure~\ref{fig:closed_loop_trajectories} shows the trajectories of the payload and the
agents in three dimensions and their projections onto the
$xz$, $zy$, and $xy$ planes. The payload trajectory deviates
slightly from the reference and exhibits low-amplitude oscillations in
some intervals. These intervals coincide with the
intervals in which one of the desired cable tensions is held at its
lower bound.

\vspace{4pt}   

\begin{figure}[h]
    \centering
    \includegraphics[height=6cm,width=\linewidth,keepaspectratio]{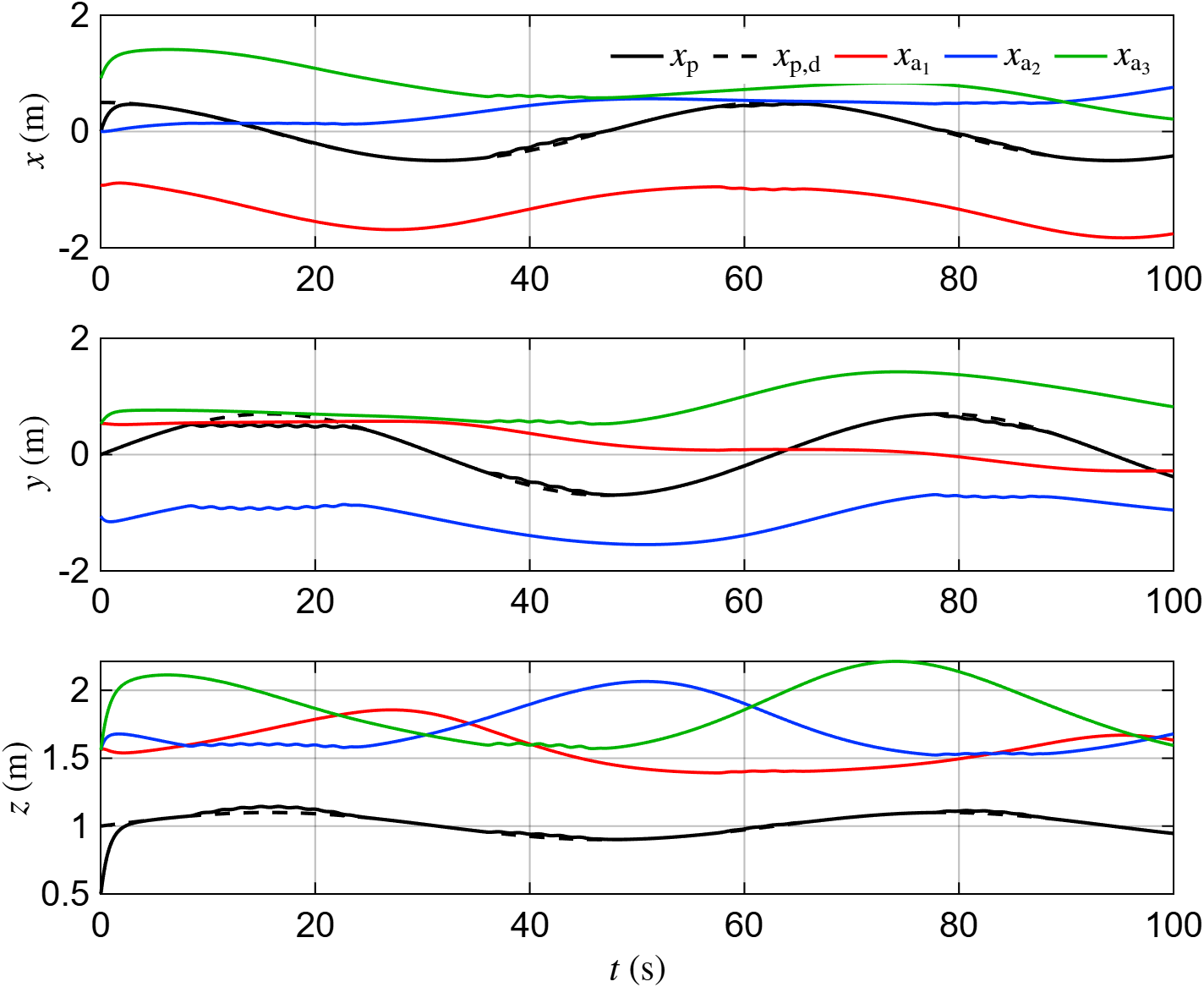}
    \includegraphics[height=6cm,width=\linewidth,keepaspectratio]{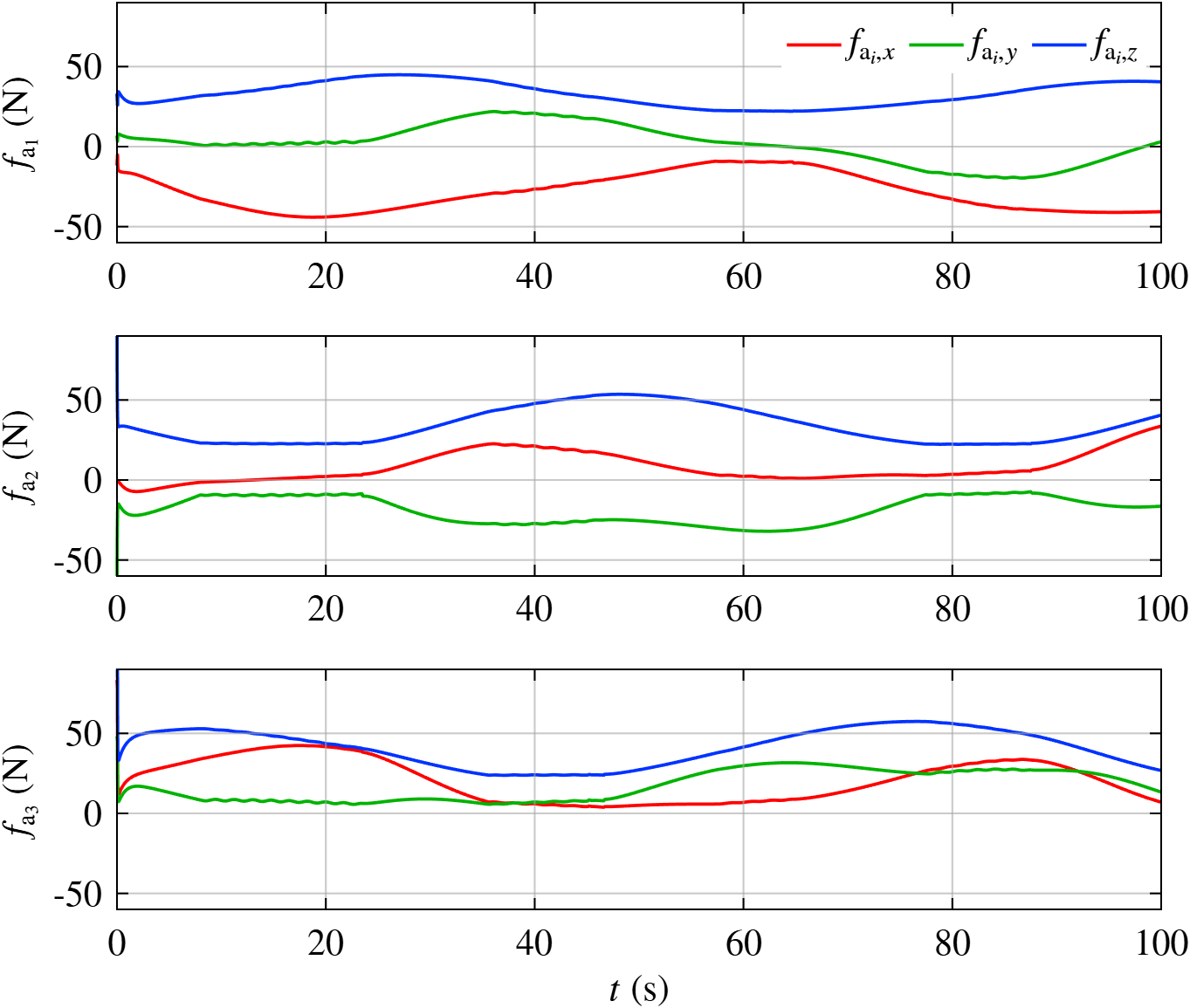}       
    \caption{Ideal force realization. Positions of the payload and the agents (left) and agent forces (right).The dashed lines show the desired payload position (left).}
    \label{fig:Position and forces}
\end{figure}

Figure~\ref{fig:Position and forces}  shows the Cartesian components of the payload and agent positions (left) and the components of the agent forces $f_{\mathrm{a}_i}$ given by \eqref{eq:tension_control_total_force} (right). In the left column, the desired payload position is shown in dashed black and the actual payload position in solid black. After the initial transient, the payload follows the desired trajectory in all three coordinates, with a nonzero tracking error $e_\mathrm{p} = x_\mathrm{p} - x_\mathrm{p,d}$ over portions of the maneuver. In the right column, the upper, middle, and lower panels correspond to forces of agents $a_1$, $a_2$, and $a_3$, respectively. The $x$, $y$, and $z$ components of each agent forces are shown in red, green, and blue. The large forces at $t = 0$ are caused by the initial position error of the payload.

\begin{figure}[h]
    \centering
    \includegraphics[width=0.5\linewidth]
    {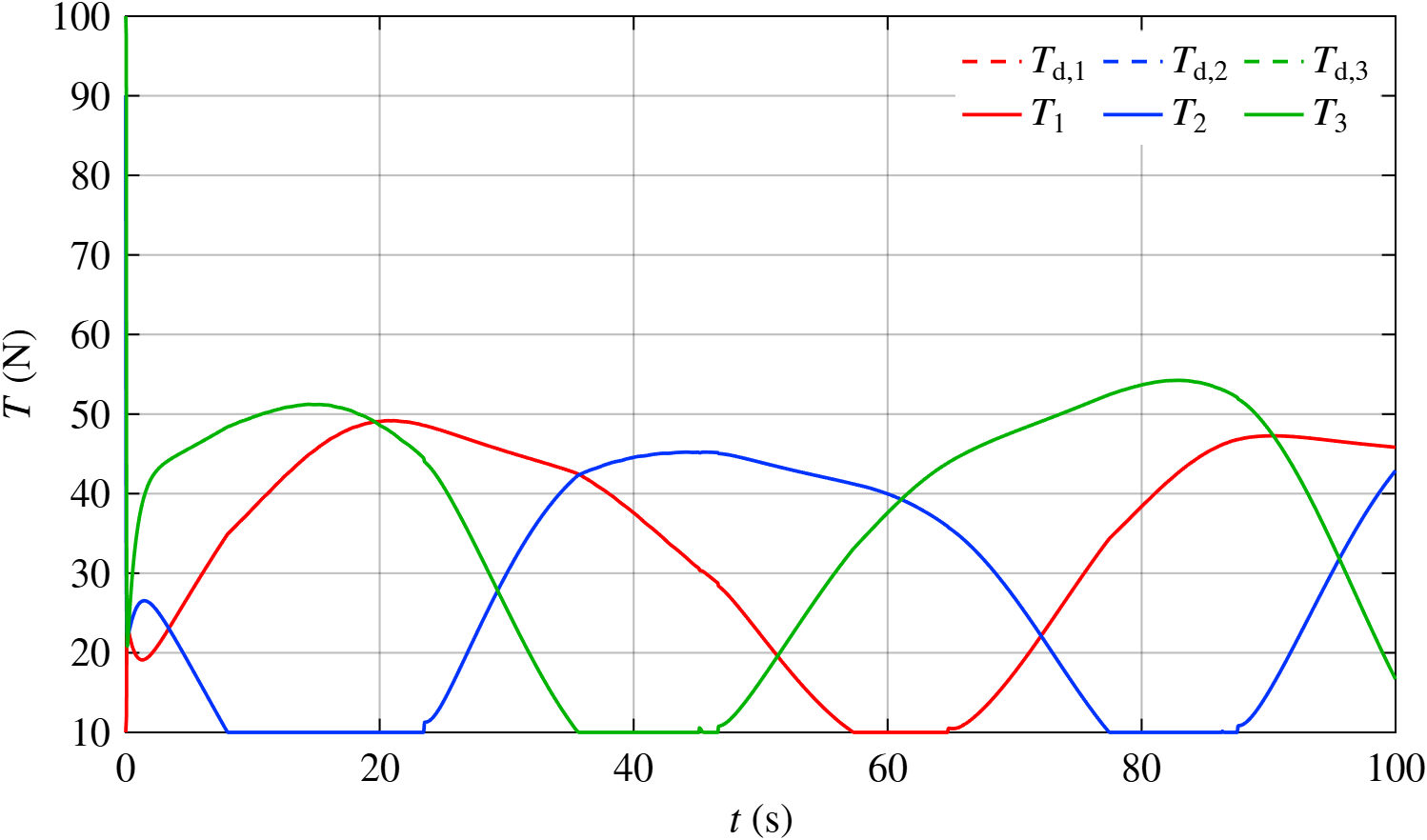}
    \caption{Ideal force realization. Desired cable tensions $T_{{\rm d},i}$ and cable tensions $T_i$ resulting from the coupled dynamics, for $i=1,2,3$. Note that the desired tensions are shown in dashed lines, and the actual cable tensions are shown in solid lines, and the two coincide.}
    \label{fig:closed_loop_tensions_ideal}
\end{figure}

Figure~\ref{fig:closed_loop_tensions_ideal} shows the desired cable tensions
$T_{{\rm d},i}$ obtained from \eqref{eq:tension_control_qp_selection} 
and the cable tensions $T_i$ given by  \eqref{eq:tension_solution} along the closed-loop solution, for $i=1,2,3$. The two coincide, as guaranteed by the force realization \eqref{eq:tension_control_explicit_agent_force}, and every tension remains at or above $T_0 = 10$~N, so the cables remain taut. At
$t = 0$, the initial position error produces a large demand in tensions.

\subsection{Decentralized Force Realization}
\label{subsec:sim_decentralized}
In the second scenario, the control force applied to each
agent is computed using the decentralized approximation
developed in
Section~\ref{subsec:decentralized_force_realization}.
The inter-agent coupling terms in the force realization
are neglected, while the outer-loop tension allocation
remains unchanged.
The resulting closed-loop response is compared with the
ideal force realization to evaluate the effect of neglecting
inter-agent coupling on payload tracking and cable tensions.

\begin{figure}[h]
    \centering
    \includegraphics[width=.5\linewidth]{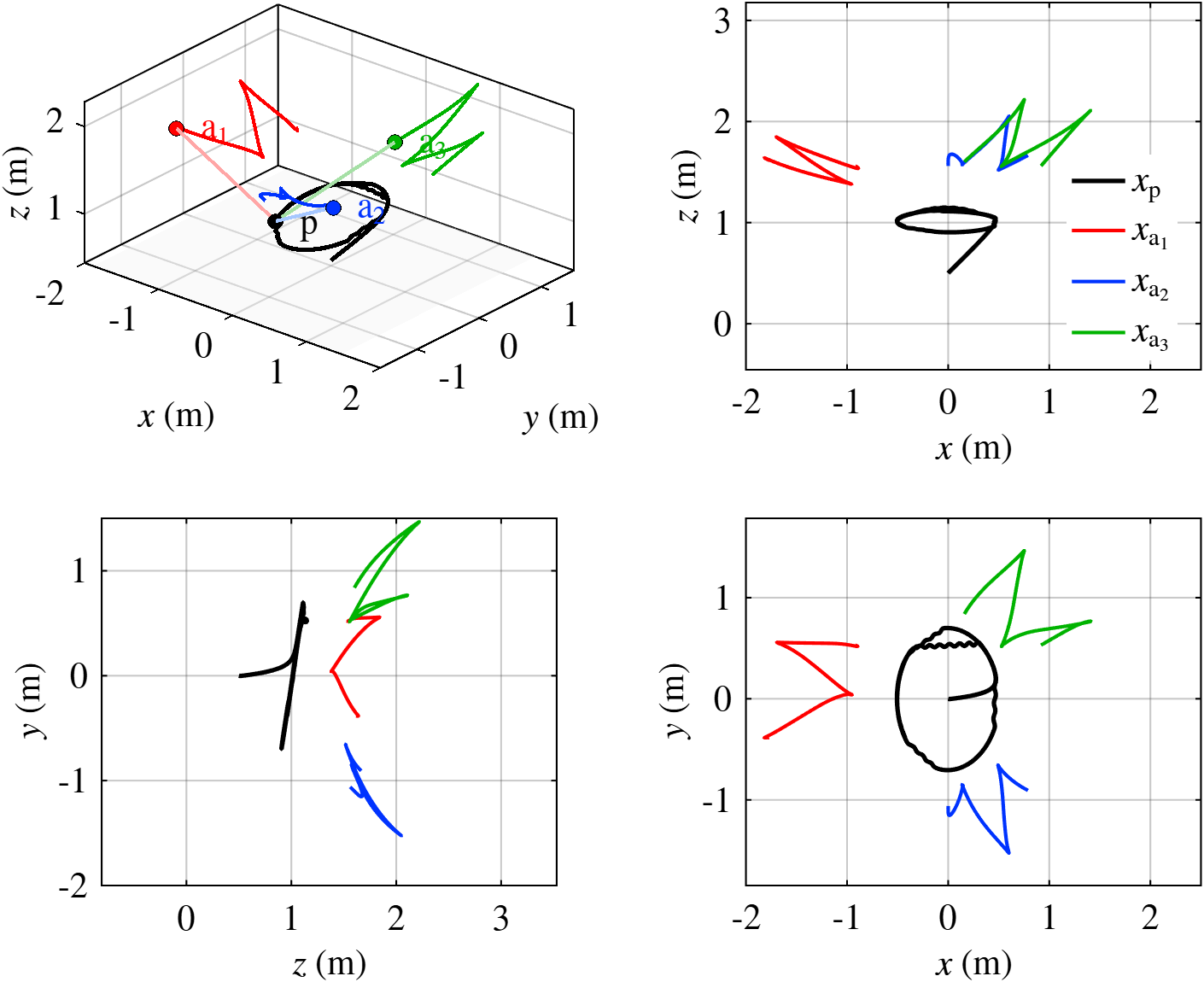}
    \caption{Decentralized force realization. Trajectories of the payload$P$ and the agents $a_1$, $a_2$, and $a_3$ in three dimensions (topleft) and their projections onto the $xz$ (top right), $zy$ (bottom left), and $xy$ (bottom right) planes. Note that the payload trajectory is shown in solid black, and the trajectories of $a_1$, $a_2$, and $a_3$ are shown in red, blue, and green, respectively.}
    \label{fig:closed_loop_trajectories_dec}
\end{figure}

Figure~\ref{fig:closed_loop_trajectories_dec} shows the trajectories of the payload and
the agents in three dimensions and their projections onto the $xz$, $zy$, and $xy$ planes. As in the ideal case, the payload follows the elliptical reference, and low-amplitude oscillations appear in the
intervals in which one of the desired cable tensions is held at its lower bound.

\begin{figure}[h]
    \centering
    \includegraphics[height=6cm,width=\linewidth,keepaspectratio]{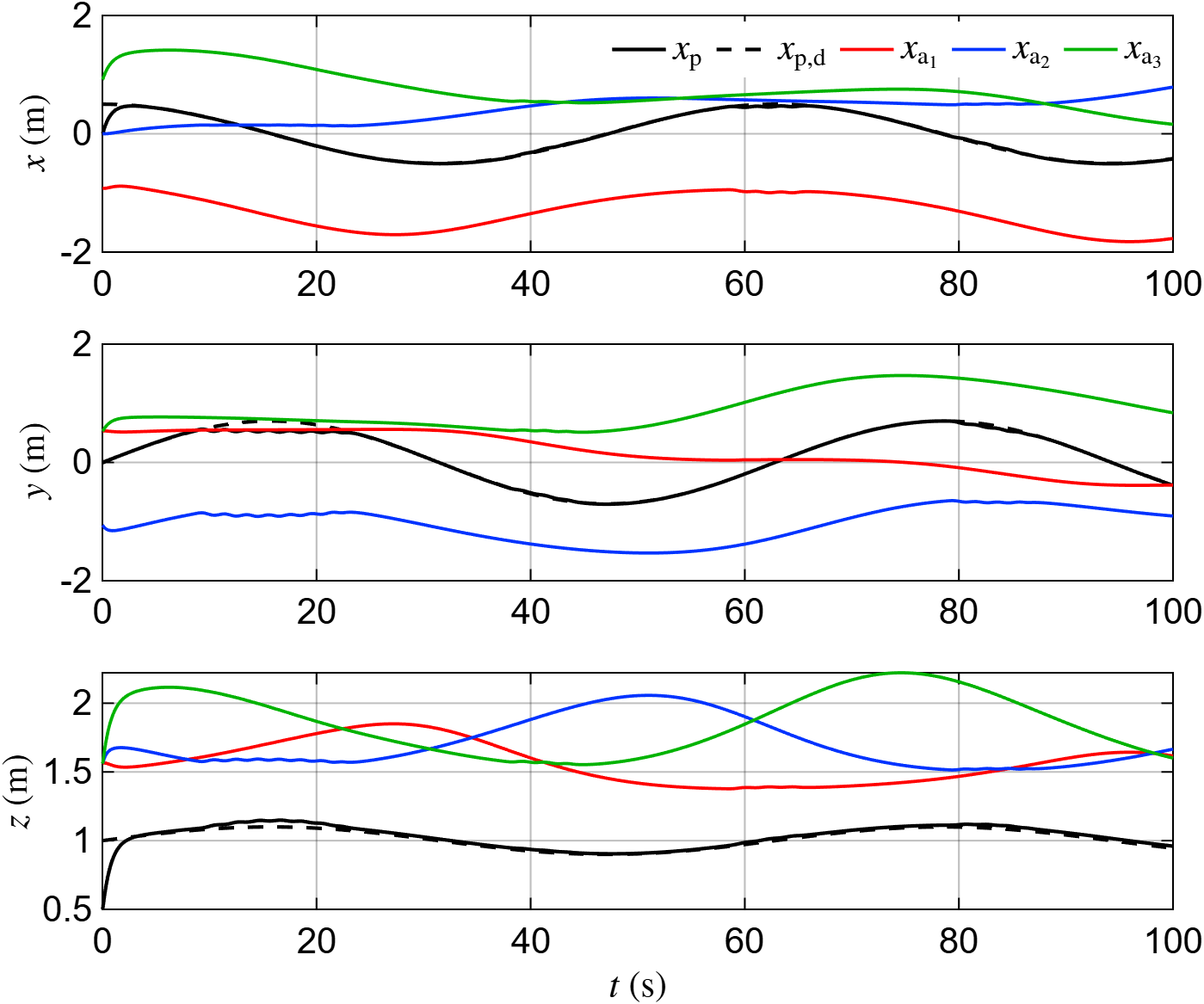}
    \includegraphics[height=6cm,width=\linewidth,keepaspectratio]{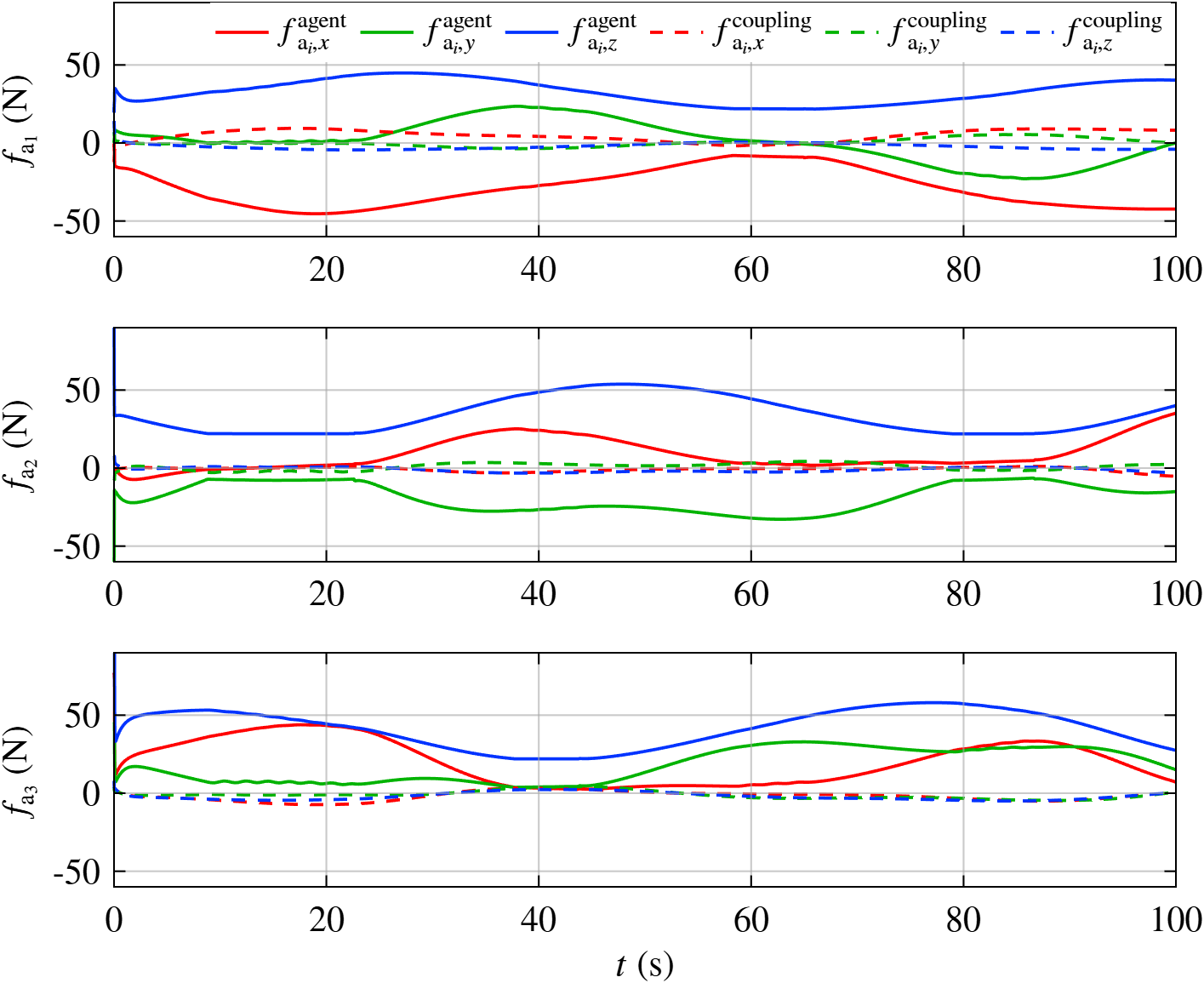}
        
        \caption{Decentralized force realization. Positions of the payload and the
        agents (left) and agent forces (right). The dashed lines show the desired
        payload position (left) and the neglected coupling forces (right).}
        \label{fig:closed_loop_force_and_position_dec}
\end{figure}

The left column of Fig.~\ref{fig:closed_loop_force_and_position_dec} shows the Cartesian components of the payload and agent positions.As in the ideal case, after the initial transient, the payload follows the desired trajectory in all three coordinates. The right column  of Fig.~\ref{fig:closed_loop_force_and_position_dec} shows the components of the agent forces $f_{\mathrm{a}_i}$ given by \eqref{eq:tension_control_total_force} with $f^\mathrm{r}_{\mathrm{a}_i} = f^\mathrm{local}_{\mathrm{a}_i}$, together with the neglected coupling forces $f^\mathrm{coupling}_{\mathrm{a}_i}$ from \eqref{eq:decentralized_force_realization_a}, shown in dashed lines. The upper, middle, and lower panels correspond to forces of agents $a_1$, $a_2$, and $a_3$, respectively, and the $x$, $y$, and $z$ components of the forces are shown in red, green, and blue. The large forces at $t = 0$ are caused by the initial position error of the payload.


\begin{figure}[h]
    \centering
    \includegraphics[width=0.45\linewidth]
    {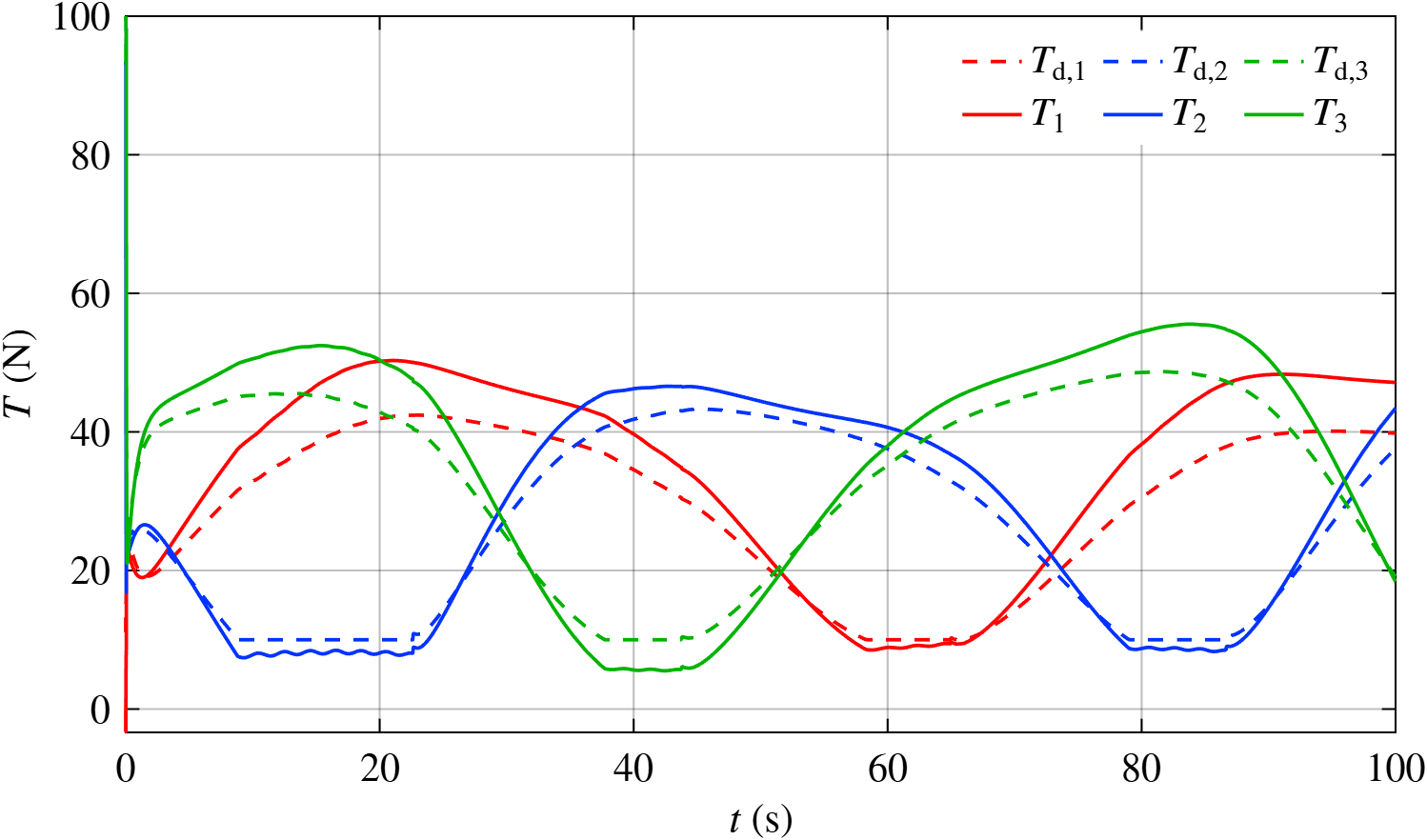}
    \caption{Decentralized force realization. Desired cable tensions $T_{{\rm d},i}$ and cable tensions $T_i$
    resulting from the coupled dynamics, for $i=1,2,3$. Note that the desired tensions
    are shown in dashed lines, and the cable tensions are shown in solid
    lines.}
    \label{fig:closed_loop_tensions_dec}
\end{figure}

Figure~\ref{fig:closed_loop_tensions_dec} shows the desired cable tensions
$T_{\rmd,i}$ and the realized cable tensions $T_i$ obtained with the
decentralized force realization.
Unlike the ideal case, the realized tensions do not equal the desired
tensions, since each agent neglects the effect of the other cables when calculating its
own tension.
All cable tensions remain positive after the initial transient, and thus
the cables remain taut.

\subsection{Effect of $T_{0}$}

In Sections~\ref{subsec:sim_ideal} and \ref{subsec:sim_decentralized}, the low-amplitude oscillations of the payload occur in the intervals in which one of the desired cable tensions is held at its lower bound.
To examine this effect, both scenarios are repeated with each entry of the lower bound $T_0$ set to $10$, $5$, $2$, and $1\,\mathrm{N}$ and all other parameters, initial conditions, and the desired trajectory \eqref{eq:simulation_reference_trajectory} are unchanged.

With ideal force realization, decreasing $T_0$ removes the oscillations associated with the active lower bound, and for $T_0=2$ and $1\,\mathrm{N}$, the oscillations are largely eliminated. Payload trajectories for the decentralized force realization also behave in the same way. However, unlike the ideal case, the payload trajectories for all values of $T_0$ remain displaced from the reference along portions of the ellipse, and this displacement does not decrease as $T_0$ decreases.

Note that a smaller $T_0$ therefore reduces the oscillations, but it also allows the cable tensions to approach zero. This effect is more significant for the decentralized force realization, in which the realized tension falls below the desired tension whenever the desired tension is held at $T_0$, as seen in Fig. \ref{fig:closed_loop_tensions_dec}. This makes the cable more likely to become slack.

\begin{figure}[h]
\centering
\includegraphics[height=6cm,width=\linewidth,keepaspectratio]{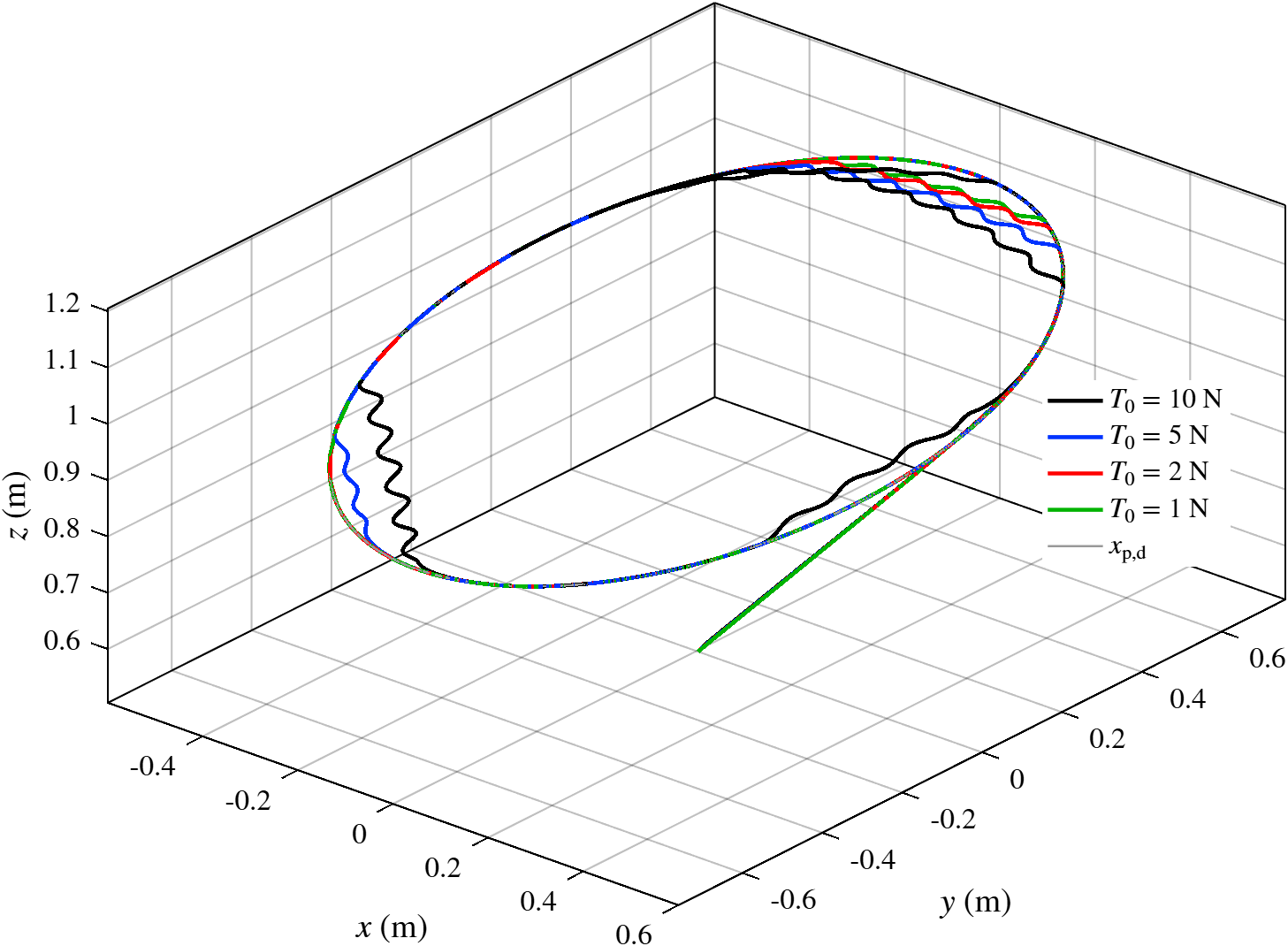}
\includegraphics[height=6cm,width=\linewidth,keepaspectratio]{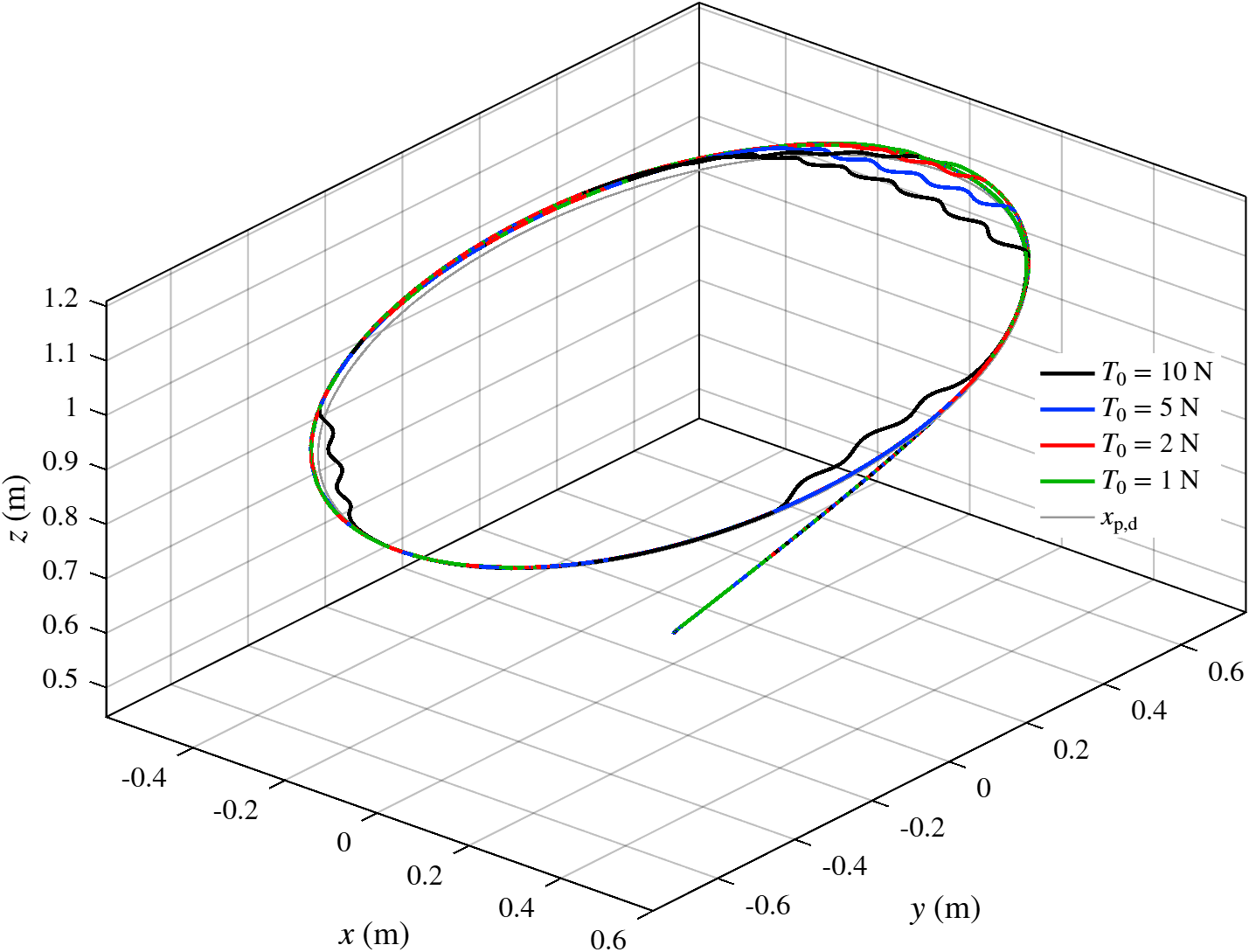}
     \caption{Payload trajectories obtained with the ideal force realization (left) and the decentralized force realization (right) for $T_0=10$, $5$, $2$, and $1\,\mathrm{N}$. Note that the trajectories for $T_0=10$, $5$, $2$, and $1\,\mathrm{N}$ are shown in black, blue, red, and green, and the desired payload trajectory $x_{\rmp,\rmd}$ is shown in gray.}
    \label{fig:closed_loop_force_ideal}
\end{figure}


\section{Limitations and Future Work}
\label{sec:limitations}
The proposed control framework is developed under several simplifying assumptions that will be addressed in future work.
First, the agents are modeled as point masses with directly realizable control forces, and the cables are assumed to be massless, inextensible, and continuously taut.
In practice, multirotor vehicles generate forces through their attitude and thrust dynamics, which introduce additional constraints on the magnitude and direction of the forces available for payload transportation.
Future work will investigate how the hierarchical controller can be extended to account for these dynamics while maintaining the desired cable tensions and payload tracking performance.

Second, the centralized force-realization controller assumes knowledge of the system parameters and the coupled payload-agent dynamics.
Errors in payload mass, agent masses, and cable geometry can therefore cause discrepancies between the desired and realized tensions.
An important direction is to investigate whether online adaptation can compensate for these uncertainties while preserving positive cable tensions and acceptable payload tracking performance.

Finally, the decentralized force-realization approximation neglects the coupling between agents and consequently does not guarantee that the realized tensions satisfy their prescribed lower bounds.
The numerical results illustrate the resulting tradeoff between payload tracking performance and the margin against cable slack.
Future work will investigate decentralized control architectures that account for or estimate the effects of inter-agent coupling using locally available measurements, with the objective of recovering tension and tracking guarantees without requiring complete system-state information.

\section{Conclusions}
\label{sec:conclusions}
This paper developed a constrained dynamical model and a hierarchical feedback controller for collaborative payload lifting using multiple agents connected to a common payload through ideal cables.
By enforcing the cable-length constraints and their acceleration compatibility conditions, an explicit relationship between the cable tensions, system state, and agent forces was obtained.
This relationship was used to develop a hierarchical control architecture in which an outer-loop controller determines the desired cable tensions through constrained optimization, while an inner-loop controller computes the agent forces required to realize those tensions.
The remaining freedom in the agent forces was used to regulate cable geometry.

A decentralized force-realization approximation was also investigated by neglecting inter-agent coupling terms in the control law.
Numerical simulations of a three-agent system demonstrated payload trajectory tracking under both the centralized and decentralized implementations.
Under the assumed model, the centralized controller realized the desired cable tensions, whereas the decentralized approximation introduced discrepancies
between desired and realized tensions and increased
payload tracking errors.
Additional simulations demonstrated the effect of the prescribed minimum cable tension on tracking performance and cable-tension margins.

The results provide a foundation for further investigation of adaptive and decentralized control of collaborative payload transportation systems under uncertain dynamics and realistic agent actuation constraints.

\section*{Acknowledgments}

AI tools were used solely to refine the organization, language, and clarity of the manuscript.
All research ideas, technical developments, analyses, and results originated from the authors.
No AI-generated scientific content was used, and the authors reviewed and verified the final manuscript.

\bibliography{bib_paper}

\end{document}